# On the Stability of the Interstellar Wind through the Solar System

**Jean-Loup Bertaux (1) and Rosine Lallement (2)**

**1.** LATMOS/IPSL, Université de Versailles Saint Quentin, INSU/CNRS, 11 Bd D' Alembert, 78200 Guyancourt, France 2. GEPI Observatoire de Paris, CNRS, Université Paris Diderot, Place Jules Janssen 92190 Meudon, France

jean-loup.bertaux@latmos.ipsl.fr

**Abstract**. As a follow-up of a recent study [1], we challenge the claim that the flow of interstellar helium through the solar system has changed substantially over the last decades. We argue that only the IBEX-Lo 2009/2010 measurements are discrepant with older consensus values. Then we show that the probability of the claimed variations of longitude and velocity are highly unlikely ($\approx 1\%$), in view of the absence of change in latitude and absence of change in the $(V, \lambda_\infty)$ relation, while random values would be expected. Finally, we report other independent studies showing the stability of Helium flow and the Hydrogen flow over the years 1996-2012, consistent with the 70's earlier determinations of the interstellar flow.

## 1. Introduction

Since the discovery of the interstellar wind (the flow of neutral H and He atoms through the solar system) with OGO-5 spacecraft in 1970 [2,3], a number of other space investigations have been used to determine the characteristics of this flow (direction, velocity, temperature). It was recently suggested [4] that the direction of interstellar helium flow has changed by $\approx 5.5°$ within the last $\approx 40$ years. At a velocity of 25 km/s, this corresponds to 210 AU, much smaller than the mean free path between collisions for an interstellar He atom ($\approx 1,000$ AU). From the point of view of the physics of rarefied gas dynamics, this is questionable, since several collisions are needed to reach a Maxwell-Boltzmann distribution which can be defined by macroscopic parameters as the temperature and the bulk velocity. Rather than discussing this aspect, several comments are now developed concerning a number of He wind direction measurements assembled in [4] to claim a secular change of the He flow direction. Within the frame of an ISSI workshop in 2003-2004 discussing UV measurements of resonance fluorescence at 58.4 nm of Helium atoms, *in-situ* He atoms and pick-up He$^+$ ion measurements, a set of Helium flow parameters was obtained that we call here "consensus value" [5]: velocity at infinity V=26. 24 ± 0.45 km/s, ecliptic longitude of downwind flow (J2000) $\lambda_\infty$=74.68 ± 0.56°, latitude $\beta_\infty$= -5.31 ± 0.28°, temperature T= 6306 ± 390 K. Figure 1 is extracted from [4], where we have added an index corresponding to each of the following comments.

1. The pick-up Ions (PUI) measured by PLASTIC instrument on board STEREO mission of $He^+$, $O^+$, $Ne^+$ are forming a crescent in the upstream direction [6] with a very large angular width (160°) inadequate for a precise determination of the longitude, and also there are transport effects which may deviate the PUI away from the trajectory of the interstellar neutrals.

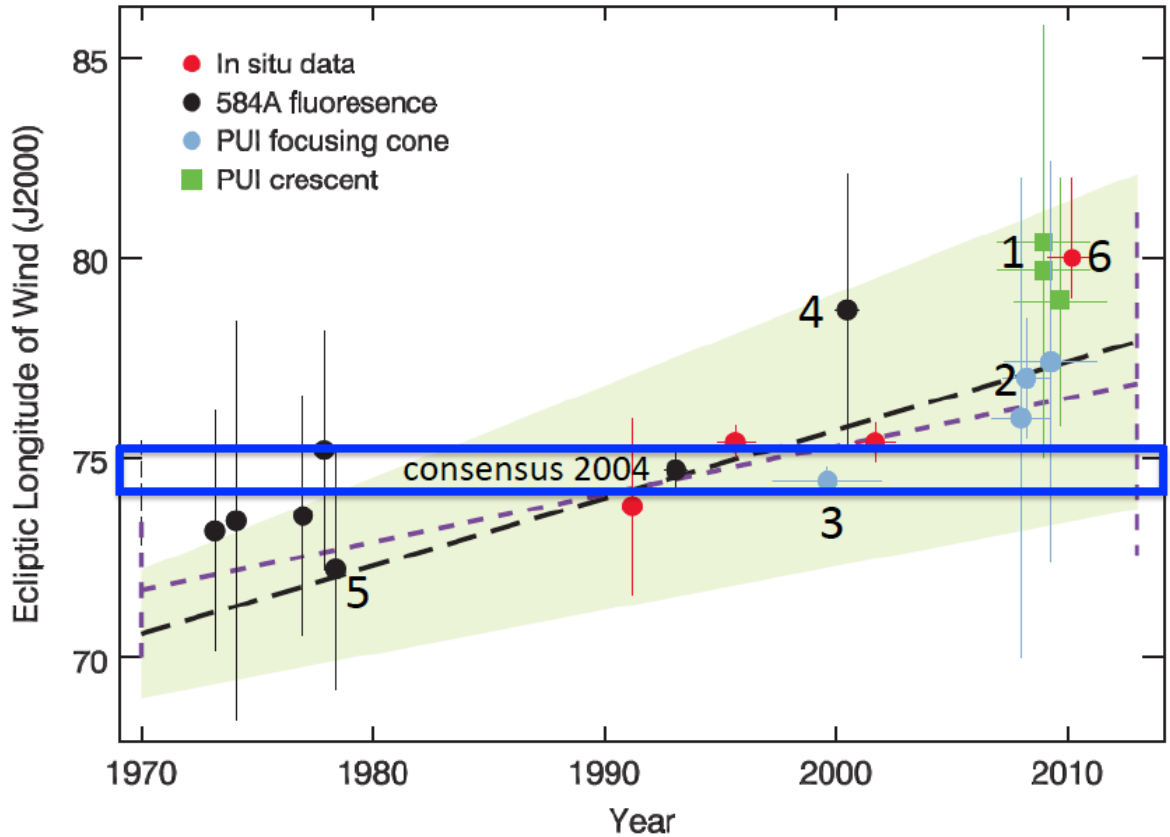

Figure 1. Adapted from [4]. This is the ecliptic longitude of the Helium downwind flow as a function of time as determined by a number of space experiments and techniques. For details, see [1,4]. Numbers from 1 to 6 are attached for reference in the discussion. The blue rectangle materializes the longitude and its error bar as discussed in [5], a consensus value combining EUV observations at 58.4 nm of He resonance, and GAS/Ulysses *in situ* measurements [15] up to 2002.

2. PLASTIC/STEREO PUI $He^+$ measured in the focusing cone [6]. While the $He^+$ *in situ* measurements [6] display a maximum at a longitude of 75.0 ± 0.3° over several years, the authors [6] use a complicated method combining by pair the Earth's orbits, with the surprising result that the He cone would be at a longitude of 77.4 ± 1.9° as plotted on figure 1. This was recently criticized in [1], from which is based the analysis of the present presentation. However, assuming that the analysis of [6] is correct, it has been demonstrated by Chalov and Fahr [7] and more recently Chalov [8] through detailed modeling that the $He^+$ maximum measured in the cone may be deflected from the He cone to larger longitudes by solar wind transport effects.

3. SWICS/ACE PUI $He^+$ measured in the focusing cone [9] in the period 1998-2002 are giving a downwind direction of $\lambda_\infty = 75.13 ± 0.33°$. It is shown in [8] through modeling that "the angular displacement of the cone in ecliptic longitude depends on the ion mean free path. When the mean free path is small (high level of turbulence), the ion velocity distribution function is close to isotropy and the displacement is small or absent at all. " This was the case for 1998-2002 conditions of SWICS/ACE, but not for PLASTIC/STEREO. All in all, the longitude of the maxima of He$^+$ should be considered as upper limits only.

4. The error bar given for the Nozomi observations [10] of He I 58.4 nm in 2001 is wrongly displayed as being ± 3.4°. We quote the paper of Nagakawa et al. [10]: « Observations suggest that the helium flow arrives from the ecliptic coordinates of (258.7 °± 3.4°), which within error bars is consistent with the primary helium flow direction derived from direct observations (e.g. Witte et al. 2004) ». There is an unfortunate typographic error in this sentence. Nagakawa et al. [10] meant: "…arrives from the ecliptic coordinates of (258.7°, + 3.4°) ». Otherwise, it would be written «…arrives from the ecliptic *longitude* of (258.7 °± 3.4°) ». Actually, there are no error bar indicated for the He flow longitude in the Nagakawa et al. paper, and the error is likely much larger than erroneously quoted by [4], probably 6° as discussed in [1].

5. This point at 72.2° is taken from the analysis of Prognoz 6 58.4 nm photometric measurements (Dalaudier et al., 1984, [11]). However, the authors [11] quote the value of 72.2° as "an eye fit preliminary determination" based on a fraction of the data, and it is clear from the whole article that it can by far not compete with the second result, based on the whole dataset and a careful data-model adjustment. The abstract of [11] contains only the value of 74.5 ± 3°, correctly corrected into 75.2 ± 3° and plotted on figure 1 by [4] to account for the change between J1950 and J2000 (+0.7° increase of ecliptic longitudes). We conclude that the use of the first value 72.2° to determine a linear time trend is not justified.

6. The only solid point away from the consensus value is the determination of the longitude in 2009-2010 from *in situ* measurements of neutral He atom from the IBEX-Lo instrument, analyzed very seriously and thoroughly in Möbius et al [12] and Bzowski et al. [13]. From essentially the same data sets, a longitude of 79.0° +3.0°, (− 3.5°) is found in [12], while a longitude of 79.2° is found in [9], but with no error bar quoted. In McComas et al [14], a combination of IBEX values is claimed to give (their Table 1) $\lambda_\infty$=79.00° and 1σ uncertainty of 0.47°, with no justification of this lower error bar. Note that this new value of 79.0° is called "consensus value" in [14] inappropriately, since it results from two IBEX studies of the same data set.

Strangely enough, the IBEX red point 6 on figure 1 is at a longitude $\lambda_\infty$=80.0° instead of the original IBEX value of 79.0°. In order to understand this suspicious upward shift of 1°, one has to read carefully a footnote in Table S1 of the Supplementary On line Material of [4]. We quote:
"This He flow direction [IBEX, 79.0°+3.0, -3.5] is further constrained to be 80.0°+2.0, -1.0 degrees by restricting the longitude range using the independent LIC temperature obtained from Sirius data (SM-S4) combined with the longitude parameter range shown in Fig. 1 of [our reference [14]]". The argumentation in [4, Supplement Online Material] is the following: the width of interstellar absorption lines to star Sirius (at a distance of 2.7 parsec) of Fe$^+$, Mg$^+$, and Ca$^+$ ions indicate a temperature of 5800 +300 -700K. When these values are fed into the IBEX relationship (longitude, T) (Fig. 1 of [14]), it converts into a longitude and its error bar which then is combined with the IBEX longitude of 79.0° +3.0, -3.5° into a new value of 80.0°+2.0, -1.0°. We disagree totally with this approach that is not consistent because the interstellar temperature parameter to Sirius is averaged over 2.67x2.06 10$^5$ =5.5 10$^5$ Astronomical Units, while the whole paper [4] seeks to demonstrate changes at a scale of 200 AU! This average value should not be combined with local values. At the same time the IBEX point is moved upward by 1°, the original error bar (+3.0°, -3.5°) is also strongly reduced without any justification. Being at the extreme boundary of the examined time and longitude intervals, this manipulation of the original data increases the weight of this point by a factor of 4, which has certainly consequences in the linear least square fit of the series of data points and its confidence interval.

In addition, there is an issue to use Sirius as a proxy of ISM local temperature, since it has multiple components and no usable D I line. There are much better determinations of the temperature of the LIC that suggest significantly higher T~7500 K [15] and according to B. Wood (private communication) the best single line of sight to represent the LIC at the solar system would probably be Eps Eri (d=3.3 pc, T=7800 ± 800 K, in [16]). If these higher temperatures were considered, it would push down the longitude instead of increasing it.

## 2. Likelihood of the claimed He changes

The IBEX-Lo longitude value is significantly different (by 4.3°) from the old consensus value. It may be either that there is a historical change, as claimed in [4], or one (or two) of the longitude determination(s) (consensus or IBEX) is (are) wrong. We show that the former situation is very unlikely. Indeed, the ecliptic latitude of the flow measured by IBEX is basically unchanged (at -4.9 ± 0.2°, point A on figure 2a), from the previous consensus value ($\beta_\infty$= -5.31 ± 0.28), while the longitude has changed (point B). The change in latitude is 0.41 ± 0.34°.

On figure 2a is illustrated the situation in the sky plane. With a motion of 4.3° in a random direction from the old consensus value (marked A) to point B, it could have moved anywhere along a circle of 4.3° radius centered on point A. The probability that the latitude of a random change is within 0.4° from 0 in the motion from A to B is 0.8°/4.3°*π= 6 %, if one considers the fact that it could also fall on the other side of A at the same latitude.

On the other hand, the IBEX show a change of the velocity of He flow at infinity, previously at V=26. 24 ± 0.45 km/s and now at 23 km/s. On figure 2b are plotted the positions of point A and B in a graph (V, $\lambda_\infty$). If the wind is changing, changes in V and $\lambda_\infty$ should be de-correlated: the new velocity V(B) could have any new value within ± 5 km/s from V(A) (this value of a velocity change ± 5 km/s is rather arbitrary, but not very different from the claimed V(A)-V(B)). However, the new couple V(B), $\lambda_\infty$(B) falls exactly on a well known curve V($\lambda_\infty$), which is the locus of all couples (V, $\lambda_\infty$) giving hyperbolic trajectories tangential to the same Earth longitude position $\lambda^E_{max}$= 130° (figure 3). Therefore, this is where an instrument like IBEX-Lo, in the configuration of rotating about the sun direction, looking at 90° from sun direction, would measure a maximum of the He flux. The old couple V(A), $\lambda_\infty$(A) is also exactly on the same locus. The probability of claimed observed change of $\lambda_\infty$ while staying on the same locus (V, $\lambda_\infty$) is the ratio of the thickness ΔV(locus)≈1 km/s of the locus to the (somewhat arbitrary) extent of the possible change of V, ≈10 km/s. The ratio and then the probability of the observed change is therefore ΔV(locus)/ 10 km/s = 0.1.

Therefore, the combined probability of observing the particular values of changes as claimed by IBEX is of the order of 0.6%, which is very unlikely. What is more likely is that one of the Helium parameters measurements (IBEX or old consensus) is wrong.

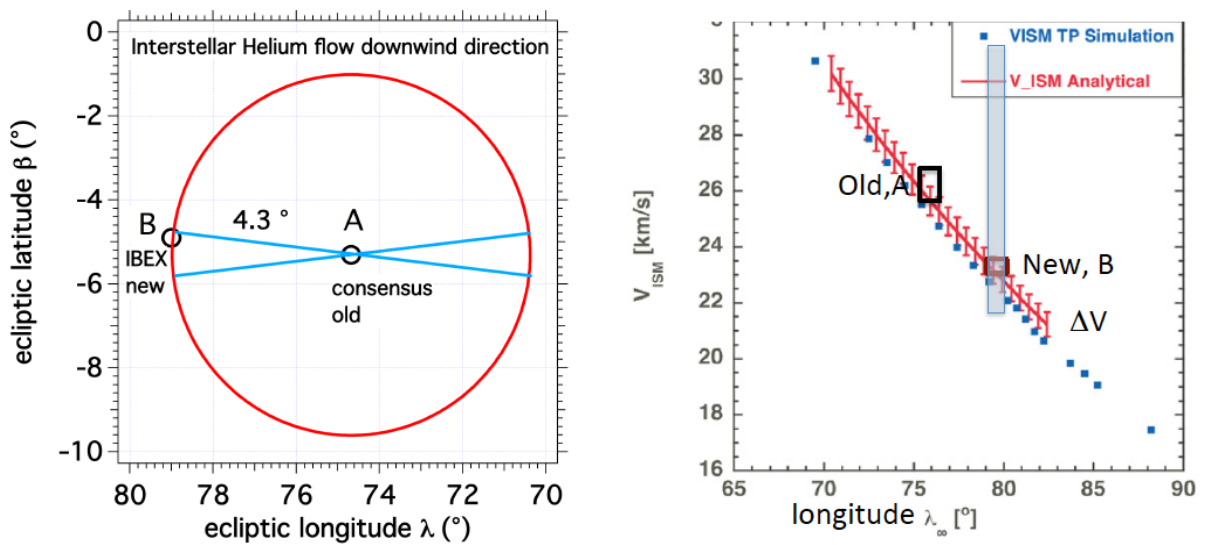

Figure 2a (left): The positions in the sky plane of He downwind direction flow, as determined earlier (A, old consensus) and IBEX, point B. The probability for a random change to stay within the portion of circle limited by the blue lines (essentially no latitude change) when there is a change of longitude of 4.3° is ≈6%. Figure 2b (right): adapted from figure 23 of [13]. There is a relationship between the downwind longitude $\lambda_\infty$ and the velocity V imposed by the Earth longitude at which is measured the maximum He flux. Blue points are the analytical approach of [12]; red bars are the model results of [13]. A marks the old consensus values and B the new IBEX value. If there were a random change of the interstellar wind parameters, there would be no reason why the B point would be on the same locus as point A. It could be anywhere along the grey rectangle.

## 3. Non-Linearity / Dead Time correction effect

It was noted by Lallement and Bertaux [1] that the two coincidences mentioned above have a common particularity. When IBEX (in terrestrial orbit) is not far from longitude position $\lambda^E_{max}$= 130°, the latitude of incoming flow is easily determined from the latitude where is measured the maximum He count rate, along one scan of the rotation around the sun pointing axis. And the particular locus (V, $\lambda_\infty$) is determined by the Earth longitude $\lambda^E_{max}$ at which the He count rate is maximum, looking at 90° from sun direction. Therefore, IBEX-Lo finds angular positions of maximum flux where they should be if the old values are valid. However, what imposes new values for the longitude and velocity is the way the signal decreases when IBEX is further away from $\lambda^E_{max}$= 130°. The counted maxima are decreasing more slowly than would be predicted by a model with the old values. An instrumental effect like a non-linearity could mimic such a behavior, if higher fluxes are producing relatively smaller counting rates, as suggested by [1]. A problem of dead time in the counting chain of IBEX-Lo do present this feature, and this idea was explored more thoroughly from modeling [1]. Actually, this effect is known to exist [12], and an estimate (not very accurate) is given "…up to 5 ms". In addition to He atoms impact on the detector, there are some electrons which participate as an additional background to this dead time. Assuming a background of 22 electrons/s as discussed in [12], a simulation was performed for various putative values of the dead time. The IBEX counting rates were taken from [13] (blue circles on figure 3, the maximum count rate was 25.3 counts s$^{-1}$, then normalized to the IBEX model, red line) and corrected for the dead time. It was found that if the dead time is of 7 ms, the maximum true counting rate (corrected from dead time) would turned to be 43 count s$^{-1}$.

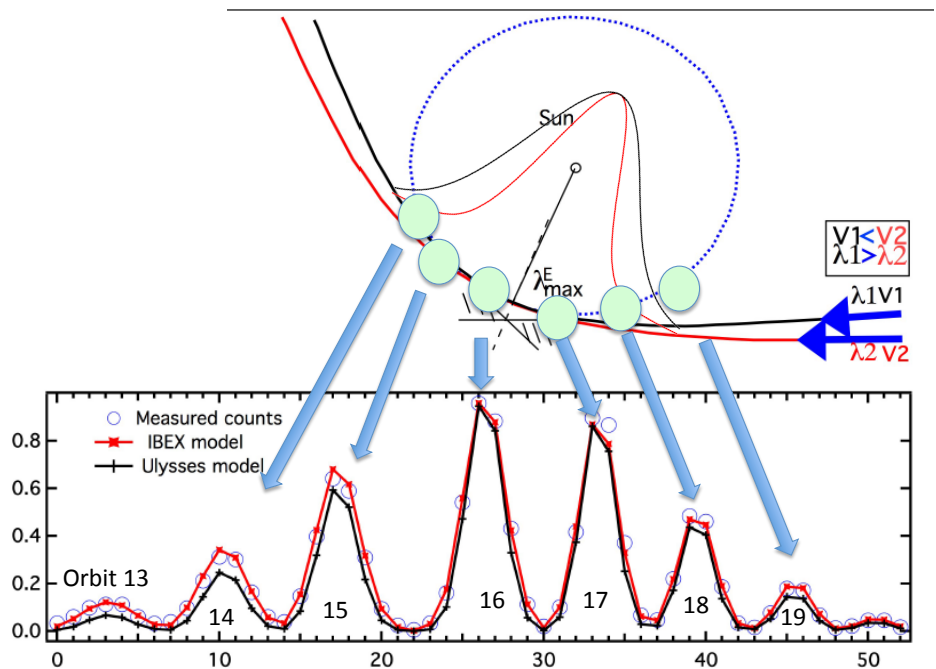

Figure 3 Top: Green circles indicate the positions on Earth's orbit where were collected the IBEX measurements, with the corresponding peak along an IBEX scan at 90° from sun plotted on bottom panel (blue circles, normalized to the best IBEX model of [13]). The data of successive orbits are concatenated for clarity. On top is illustrated that a family of hyperbolic trajectories are tangent to the Earth's orbit at the same point, giving the relationship (V, $\lambda_\infty$) of figure 2b. Bottom: the black curve (taken from [13]) is the Ulysses model [17] (basically the old consensus model). All data and model are normalized to 1 at the maximum. This Ulysses model is distinct from the IBEX model in that successive peaks are decreasing faster with distance from the maximum (*adapted from* [1]).

On figure 4, the Ulysses model (also labelled Witte et al. model [17]), which relative values were taken from [13], is normalized to this maximum counting rate. It is clear that now the corrected counting rates follows much better the overall behaviour of the model computed by [13] from Witte et al parameters. Indeed, when varying the value of the dead time, 7 ms was the value giving the best fit to this model; at bottom of Figure 4 is plotted the difference model–data (residual), showing fluctuations, but also a trend that suggests that a slightly lower longitude could significantly improve the data-model agreement.

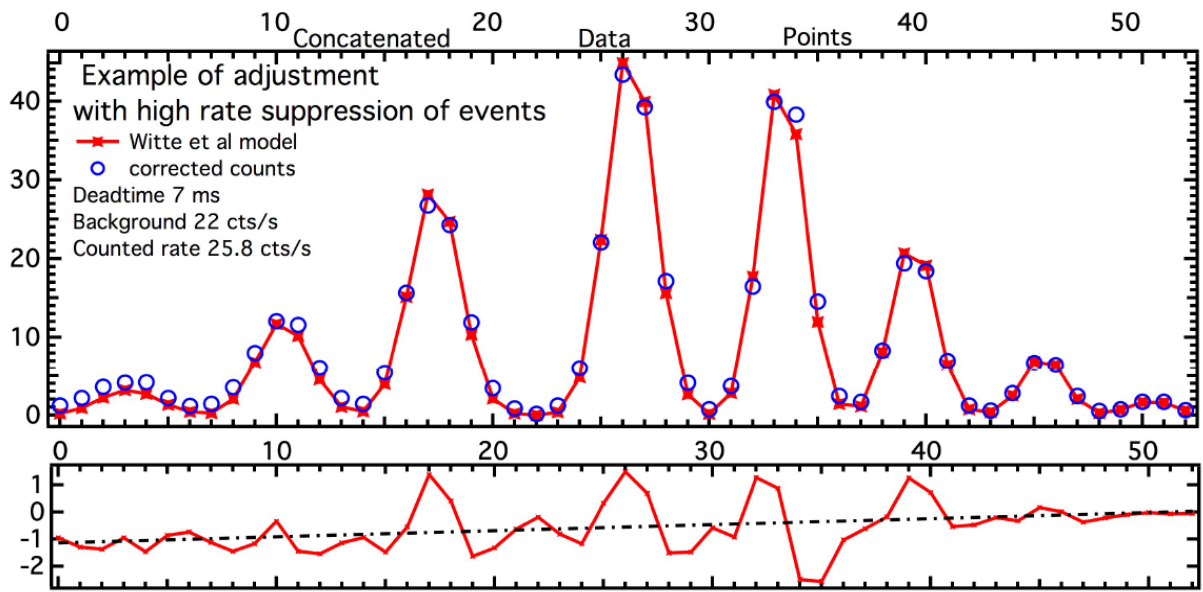

Figure 4 Blue circles: the IBEX counting rates were corrected assuming a dead-time of 7 ms, a maximum count rate of 25.8 counts/s due to He atoms and an electron background of 22 counts/s. Units are counts/s. The red curve labeled Witte et al is the same as the black curve of figure 3, labeled Ulysses with parameters from [17]. The sum of squared residuals after normalization is only slightly higher than for the [13] parameters. *Bottom*: a small asymmetry is found in all residuals, that we attribute to the fact that the Witte et al longitude is slightly too high compared to the actual flow longitude (*adapted from* [1]).

## 4. The spade of Damocles
There are a number of instruments and observations that have provided, or will provide, indications about the changes of the interstellar wind direction, or its stability. Clearly, if the Helium flow has changed, the Hydrogen flow should also have changed.

*4.1. High spectral Resolution of interplanetary Hydrogen emission with HST and SWAN.*

High spectral resolution of the interplanetary Lyman-α emission give also an information on the velocity of the H interstellar flow through the solar system. Some data coming from HST and from SWAN instrument on board SOHO (using the absorption hydrogen cell [18,19,20,21,22]) were recently [23] compared to model predictions with two different sets of parameters, labelled "Ulysses" and "IBEX" (figure 5). Along their travel through the solar system, H atoms are suffering ionization (which kills more slow atoms than fast atoms) and Lyman-α radiation pressure. The observed velocity (measured by a Doppler shift) of the H Lα line is different from the H velocity at infinity. Both effects are modulated by the solar cycle, since ionization and solar Lα both depend on the solar cycle. These effects are small at solar cycle minima (the observed velocity is near the velocity at infinity). It is clear on figure 5 that both SWAN data and HST data are more consistent with the Ulysses parameters than with the IBEX parameters. In particular, the velocity at infinity is greater than the value of 23 km/s invoked by [14] to claim the absence of bow shock outside the heliosphere. This putative bow shock is back again.

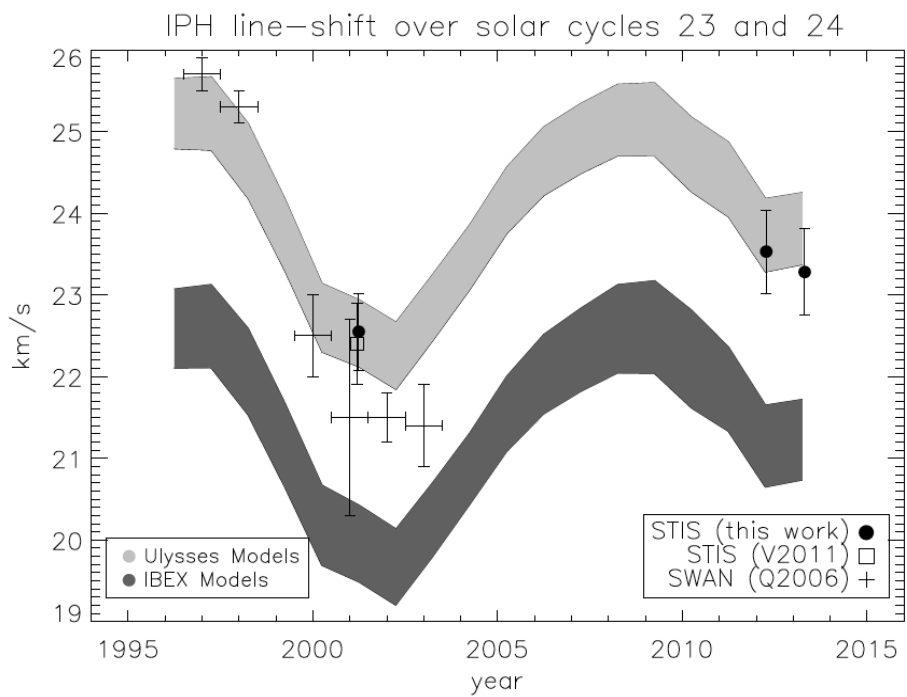

Figure 5. Taken from [21]. The observed Doppler shift of the interplanetary H Lα line was measured by HST and SWAN/SOHO several times. They are compared to a model which takes into account the effect of the solar cycle (radiation pressure, ionization) with the two sets of parameters, IBEX or Ulysses. Clearly the Ulysses parameters are favored.

*4.2. SWAN photometric Lα data in the ecliptic.*

In the solar system, the S plane, perpendicular to the ecliptic (Fig.6), containing the sun and the interstellar wind vector is a symmetry plane for H and He distributions. Therefore, particular sets of photometric data may be used to determine accurately the position of this plane. For instance, looking at right angle from the sun, in the ecliptic, but forward (F) or backward (B) (w.r.t. the Earth's motion), the intensities should be equal when the Earth is in upwind position (U) or in the downwind position (D). One complication is that the illuminating sun in Lα is not spherically symmetric because of spots, which introduces some noise. It remains to be seen if the direction is stable, and if this photometric value is consistent with the H upwind determination from SWAN Hydrogen cell absorption measurements, sensitive to the velocity of H atoms, analyzed by Lallement et al. [20].

*4.3. SWAN photometric Lα data to the North pole.*

Another set of SWAN intensities have been analyzed, all in the direction of the North ecliptic pole [24]. They are displayed on figure 7. Obviously, the intensity is maximum at position (U), and minimum at position (D). There is a strong modulation associated to the solar cycle, with less intensity at solar max, because of the stronger Lα radiation pressure and increased ionization which increase the size of the ionization cavity carved into the H flow. Figure 7 shows that intensity variations cannot be simulated by a stationary model H distribution. A much better fit is obtained with a complete 3D, time-dependent model [25]. This data set is yet to be analyzed with the prospect of determining the position of U and D. One advantage of this viewing geometry is that the intensities to the North ecliptic pole (or the South) are less sensitive to variations of the solar illumination, seeing (more or less) a complete North (or South) solar hemisphere).

*4.4 Messenger UV spectrometer*

Interplanetary Lα photometric data collected by the UV spectrometer instrument on board Messenger in the ecliptic plane are presently analyzed (Quémerais, private communication) to search for the incoming longitude of the hydrogen flow.

## Lyman α symmetry for special directions

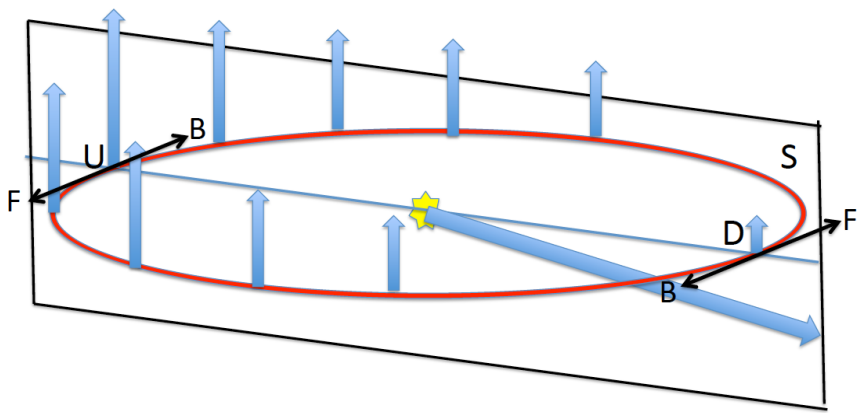

Figure 6. The S plane, perpendicular to the ecliptic containing the sun and the interstellar wind vector is a symmetry plane for H and He distributions. Looking at right angle from the sun, in the ecliptic, but forward (F) or backward (B) (w.r.t. the Earth's motion), the Lα intensities should be equal when the Earth is in upwind position (U) or in the downwind position (D). Also, when looking always to the North (or South) pole, the Lα intensity (mimicked by the length of blue arrows) should be maximum at U and minimum at D, because of the progressive destruction of H atoms by solar ionization along the flow.

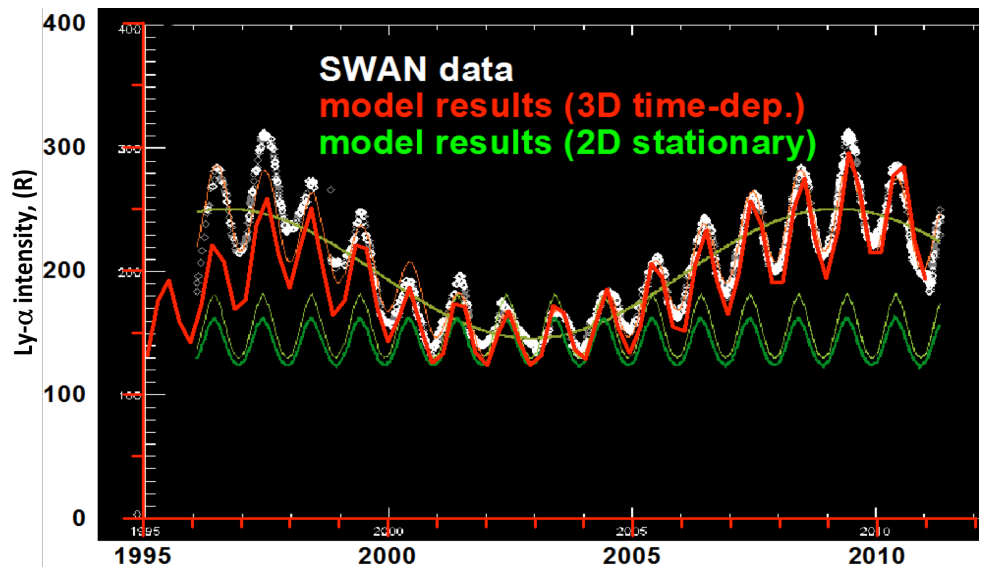

Figure 7. Evolution of the Lα intensity in the direction of the North Ecliptic Pole from SWAN/SOHO over 17 years (white diamonds). There is a solar cycle influence which is not accounted for in the 2D stationary models (green), while it is well reproduced by a time dependent/3D model (from O. Katushkina, [24]).

*4.4 Gas/Ulysses Helium flow reanalysis.*

In view of the IBEX Helium interpretations [12,13,14], three different teams have re-analyzed the Gas/Ulysses *in-situ* data, including now the 2007 data set which was not existing in the earlier analysis [15]. Wood [24] (this conference) concluded that his re-analysis was giving results similar to the original analysis. Katushkina et al. [27] did a similar work. Finally, Bzowski et al. [28] made also a re-analysis, using the same model as was used in [13] and the same fitting techniques for the analysis of IBEX helium data. They analyzed three epochs 1996, 2002, 2007 (only the fast North South orbital scan of Ulysses provide useful measurements). They find as best fit parameters for the upwind direction λ= 255.3°, β= 6° (J2000), V = 26.0 km/s, T = 7500 K. The direction and velocity are in good agreement with older findings (our consensus values) [5] while the temperature is somewhat higher.

*4.5 Angular distance between the IBEX ribbon center and the Hydrogen Deviation Plane (HDP).*

Perhaps the most spectacular IBEX result is the discovery of the so-called "ribbon", the locus on the sky from where are coming most of high energy ENAs. This structure is forming a (sometimes incomplete) circular structure around one point, the center of the ribbon. The position on the sky of this center is [29]: ecliptic longitude 219.2 ± 1.3°, latitude: + 40 ± 2.3°. On the other hand, it was discovered [22] that the incoming of Hydrogen atoms was deviated from the incoming of He atoms, because some of the interstellar $H^+$ are deviated by the heliopause obstacle, kinked by the interstellar magnetic field, and then produce H neutrals by charge exchange. Therefore, the plane containing both H and He vectors was dubbed the Hydrogen Deviation Plane (HDP), containing the interstellar magnetic field outside of the heliopause. The normal to this plane re-determined in 2010 [30] is at λ= -10.6°, β= 37.5° (ecliptic J2000), and the angle between this normal and the direction of ribbon center is 90.0°: the ribbon center is right on the HDP as determined in 2010 [30]. If the new IBEX value of He direction is taken, it gives another normal to HDP: λ= -3.2°, β= +57.5°, making an angle of 76° with the ribbon center, which would then be at 14 ° from this new HDP, a substantial deviation from the HDP.

## 5. Conclusions

The IBEX helium direction and velocity [12,13,14] are in marked discrepancy with earlier determinations of the Helium flow, either from optical measurements of He 58.4 nm resonance, or from Gas/Ulysses [17, 26, 27, 28] *in-situ* measurements. This discrepancy was interpreted as the result of a secular variation of the interstellar wind direction [4], an interpretation vigorously challenged [1]. Here we show first that the claimed variations of longitude and velocity are highly unlikely (with a probability <1%), in view of the absence of change both in latitude and in the coupling (V, $\lambda_\infty$). Therefore either the IBEX value, or the old one is true. We have then proposed an instrumental effect for IBEX-Lo sensor that could bias the fit to a model, the inability to count all He events when their rate is too high. As reported in [1], the IBEX-Lo data could be as well fitted by the old parameters, when corrected for a dead time of 7 ms, not far from a quoted possible value of 5 ms [12]. Other instrumental effects could also be investigated, as for instance the detection efficiency of Helium atoms as a function of their kinetic energy, which is assumed to be constant in the IBEX analyses [12,13].

In parallel, several independent new analyses or re-analyses of He and Hydrogen flow indicate a long term stability of the direction and velocity of the interstellar flow through the solar system. Additional analyses of other data sets are expected in the near future, telling on whose head will fall the spade of Damocles.

## 6. Acknowledgements

We thank Gary Zank for the invitation to the 13[th] Annual International Astrophysics conference in Myrtle Beach, during March 2014. We also wish to thank Dimitra Koutroumpa, Eric Quémerais, and Olga Katushkina for communication of their most recent L$\alpha$ results on SWAN. We wish to thank the referee for useful comments and Brian Wood for pointing to astronomical measurements of LIC temperature. This work was supported by CNES (Centre National des Etudes Spatiales) and CNRS (Centre National de la Recherche Scientifique) in France.## References

[1] Lallement R. and Bertaux J.L. 2014 A & A, 565, 41
[2] Bertaux J.L., and Blamont J.E. 1971 A & A , 11, 200
[3] Thomas G.E., and Krassa R.F. 1971 A & A , 11, 218
[4] Frisch P.C., Bzowski M., Livadiotis G., et al. 2013, Science, 341, 1080
[5] Möbius E., Bzowski M., Chalov S., et al. 2004, A&A, 426, 897
[6] Drews C. et al. 2012, J. Geophys. Res.,117, A09106
[7] ChalovS.V. and Fahr H. J. 2006, Astron. Lett., 32, 487
[8] Chalov S. 2014, A & A, in print
[9] Gloeckler G., Möbius E., Geiss J. et al. 2004, A&A, 426, 845
[10] Nakagawa H, et al., 2008, A&A , 491, 29
[11] Dalaudier F., Bertaux J. L., Kurt V. G., & Mironova E. N. 1984, A&A 134, 171
[12] Möbius E., Bochsler P., Bzowski M., et al. 2012, Ap.J. supp.series
[13] Bzowski, M. Kubiak M. A., Möbius E., et al. 2012, Ap.J. supp.series
[14] McComas D.J. et al. 2012, Science, 336, p.1291
[15] Redfield, S. & Linsky, J.L., 2008, Ap.J. 673, pp. 283-314.
[16] Dring,A.R., et al., 1997, Ap.J. 488, pp. 760-775
[17] Witte, M. 2004, A&A, 426, 835
[18] Bertaux J.L., Kyrölä E., Quémerais E., Lallement R. et al. 1995 Solar Physics, Volume 162, pp. 403-439
[19] Bertaux J.L., Quémerais E., Lallement R., Kyrölä E., Schmidt W. 1997 Solar Physics, Volume 175, Issue 2, pp.737-770
[20] Costa J. et al. 1999 A&A, 349, 66010